\documentclass[nofootinbib,twocolumn,showpacs,showkeys]{revtex4}
\usepackage{graphicx,color}
\usepackage{amssymb}
\usepackage{amsmath}
\usepackage{bm}
\usepackage[normalem]{ulem}
\usepackage{hyperref}

\begin{document}

\title{Neutron Stars in Rastall Gravity}

\author{A. M. Oliveira$^1$}\email{adriano.oliveira@ifes.edu.br}
\author{H. E. S. Velten$^{2,3}$}\email{velten@pq.cnpq.br}
\author{J. C. Fabris$^2$ }\email{fabris@pq.cnpq.br}
\author{L. Casarini$^2$}\email{casarini.astro@gmail.com}

\affiliation{$^1$Instituto Federal do Esp\'irito Santo (IFES), Guarapari, Brazil}
\affiliation{$^2$Universidade Federal do Esp\'{\i}rito Santo (UFES), Vit\'oria, Brazil}
\affiliation{$^3$CPT, Aix Marseille Universit\'e, UMR 7332, 13288 Marseille,  France}

\begin{abstract}
We calculate static and spherically symmetric solutions for the Rastall modification of gravity to describe Neutron Stars (NS). 
The key feature of the Rastall gravity is the non-conservation of the energy-momentum tensor proportionally to the space-time curvature. 
Using realistic equations of state for the NS interior we place a conservative bound on the non-GR behaviour of the Rastall theory which should be {$\lesssim 1\%$} level. This work presents the more stringent constraints on the deviations of GR caused by the Rastall proposal.
\end{abstract}

\maketitle

\section{Introduction}

Einstein's general relativity (GR) is a robust theory for explaining the gravitational phenomena. It operates by relating directly the curvature of space-time to its energy and momentum content by the so called field equations. Importantly, its predictions for solar system experiments and binary pulsars systems are in perfect agreement with the observational data. On larger (cosmological) scales we can keep GR as our standard gravitational theory by evoking extra unknown components like dark matter (DM) and dark energy (DE).
 
Alternative gravity theories appear usually when one allows for extra degrees of freedom or desires to abandon the concept of DM or DE components. 
Many alternative theories have been proposed as scalar-tensor theories (being Brans-Dicke\cite{Brans:1961zl} a particular case of this class), $f(R)$ theories \cite{Felicelrr2010} and others \cite{Clifton:2012gd}. Such theories can be tested with great precision using solar system tests and binary pulsars. For the latter, see \cite{wex}.

In particular, Rastall Gravity can be seen as a generalization of General Relativity.
In 1972 P. Rastall \cite{Rastall:1972ly} proposed an adjustment to Einstein's equation which causes a violation of the usual conservation law, making the covariant divergence of stress-energy tensor proportional the covariant divergence of the curvature scalar, i.e., $T^{\mu}_{\nu\, ;\mu} \propto  R_{;\nu}$.  Therefore when the local/global geometry is flat we recover Einstein's field equations. However, it seems non-trivial to explain the nature of such new source. This can be phenomenologically viewed as the appearance of quantum effects in the classical context, as discussed in \cite{Fabris:2014fk}. Interestingly, the issue of non-conservation of $T^{\mu\nu}$ is a feature also found for instance in diffusion models \cite{diffusion} and its connection to modified gravity theories has been discussed in literature in Refs.\cite{Koivisto, Mina}.  

Since the energy-momentum tensor is not conserved anymore, there is no variational principle in the context of the Riemannian geometry. However, some structures similar to that of the Rastall's theory may be found in the context of Weyl geometry \cite{romero}. External fields in the action may lead also to essentially the same structure as in Rastall's theory (for a discussion of external fields in gravity theories, see for example \cite{chauvineau}). 

Cosmological scenarios based on the Rastall proposal (Rastall Cosmology) may be degenerate with the $\Lambda$CDM at background and at first order levels, i.e., a viable Rastall cosmology can be constructed \cite{cosmo}. On the other hand, astrophysical consequences of Rastall proposal have not yet been widely considered.

Apart from the cosmological solutions, gravitational theories must also provide other solutions for stellar and black hole configurations and gravitational wave emission. Concerning the former, the widely known Tolman-Oppeinheimer-Volkoff (TOV) equation allows us to determine the GR static and spherical equilibrium configuration of the stellar interior \cite{Oppenheimer:1939qf, Tolman:1939dn}.
   
Stars generate energies by nuclear reactions within the stellar interior and gravitational forces are supported via radiative convection.
This balance can be studied with Newtonian mechanics, via the Lane-Emden equation, as far as {$p \ll \rho$ (in $c=1$ units)}.
Even white dwarfs which have the gravitational collapse supported by the electron degenerescence, maximal mass $\sim 1.4 M_\odot$ and a radius of order of hundreds of kilometers, are well described by the classical theory \cite{chandrasekhar2012introduction}.

Compact stars have {$p\approx \rho$} where the density is approximately the nuclear saturation density. Then the use of the relativistic theory, as given by the TOV equations, is mandatory. The effects of pressure can not be neglected and it is not viable to use of Newtonian theory to describe this structure (see \cite{Oliveira:2014jk} for an alternative viewpoint).

Since the essence of the Rastall gravity is associated to high curvature environments, neutron stars seem to be promising laboratories for testing it. Our main intention in this work is the establish bounds to deviations to GR caused by Rastall's proposal.

In the next section we obtain the Rastall version of the TOV equations. Numerical results of such new equations for the mass-radius diagram adopting realistic equations of state for neutron stars are shown in section \ref{sec:numsol}. Finally we conclude in last section.  

\section{Static and Spherical solutions}

The description of stellar objects is simplified when we consider that matter is spherically symmetric distributed in a static geometry. The resulting spacetime is described by  
\begin{eqnarray}
ds^2 = B(r)dt^2-A(r)dr^2-r^2(d\theta^2+\sin^2\theta d\phi^2).
\label{eq:metrica}
\end{eqnarray}
The matter content of an ideal fluid, neglecting the magnetic field and the rotational frequency, is contained in the energy-momentum tensor  
\begin{eqnarray}
T_{\mu\nu} &=& (\rho + p) u_\mu u_\nu - p g_{\mu\nu},
\end{eqnarray}
where $u_\alpha$ is the 4-velocity (with $u_\mu u^{\mu} = 1$), $g_{\mu\nu}$ is the component from metric tensor.

An important step in deriving the standard TOV equations is the conservation law,
\begin{eqnarray}
T^{\mu\nu}_{;\mu}=0,
\label{eq:conservacao}
\end{eqnarray}
which is concomitant with the use of Einstein's equations,\begin{eqnarray}
R_{\mu\nu} - \frac{1}{2}g_{\mu\nu} R = 8\pi G T_{\mu\nu},
\label{eq:Einstein}
\end{eqnarray}
where $R_{\mu\nu}$ is the Ricci tensor and $R$ is the curvature scalar (trace of tensor $R_{\mu\nu}$). After following these equations for the geometry defined in (\ref{eq:metrica}) we find then the TOV equation 
\begin{eqnarray}
\frac{dp}{dr} &=& -\frac{G M \rho}{r^2}\frac{\left(1+\frac{p}{\rho}\right)\left(1+\frac{4\pi r^3 p}{M}\right)}{1 - \frac{2GM}{r}}\quad,
\label{eq:TOV}
\end{eqnarray}
where on should keep in mind that
\begin{eqnarray}
M=\int_0^r 4\pi \rho r'^2 dr',
\label{eq:massa}
\end{eqnarray}   
is a definition used during the derivation of the TOV equation and is automatically associated to the mass contained within $r$. See \cite{weinberg1972gravitation, glendenning2000compact} for a revision of this calculation.

Let us now redo the same steps in finding (\ref{eq:TOV}), but using however the Rastall gravity. The conservation law will be
\begin{eqnarray}
T^{\mu\nu}_{;\mu} = \frac{1-\lambda}{16\pi G}R^{;\nu}. 
\label{eq:Rconservacao}
\end{eqnarray}
where the GR framework is recovered by setting $\lambda=1$.

One can recast the field equation (\ref{eq:Einstein}) according to the Rastall's proposal \cite{Rastall:1972ly} as
\begin{eqnarray}
R_{\mu\nu} - \frac{\lambda}{2}g_{\mu\nu} R = 8\pi G T_{\mu\nu}. 
\label{eq:E-R}
\end{eqnarray}

The Rastall version of the TOV equation, we call it RTOV from now on, can be written in a similar form to the original TOV equation as
\begin{eqnarray}
\frac{d\tilde{p}}{dr} = -\frac{G \tilde{M} \tilde{\rho}}{r^2}\frac{\left(1+\frac{\tilde{p}}{\tilde{\rho}}\right)\left(1+\frac{4\pi r^3 \tilde{p}}{\tilde{M}}\right)}{1 - \frac{2G\tilde{M}}{r}},
\label{eq:TOVR}
\end{eqnarray}
and 
\begin{eqnarray}
\frac{d\tilde{M}}{dr}&=&4\pi \tilde{\rho} r^2,
\label{eq:massaR}
\end{eqnarray}
where new effective quantities were defined using the superscript ``\,$\tilde{}$\,''. The new quantities $\tilde{p}$ and $\tilde{\rho}$ are related to $p$ and $\rho$ by
\begin{eqnarray}
\tilde{\rho}&=& \alpha_1 \rho + 3\alpha_2 p,\nonumber\\
\tilde{p}&=& \alpha_2 \rho + \alpha_3 p,
\label{rhoptilde}
\end{eqnarray}
where 
\begin{eqnarray}
\alpha_1 = \frac{2+3\eta}{2+4\eta};\quad\alpha_2 =\frac{\eta}{2+4\eta},\quad\alpha_3 =\frac{2+\eta}{2+4\eta}.
\label{alphas}
\end{eqnarray}
Here we assumed that $\lambda = 1+\eta$ and $\eta$ is the new term which parameterizes deviations from General Relativity.

The signs of $\alpha_1, \alpha_2$ and $\alpha_3$ are related to the stellar stability by equation (\ref{eq:TOVR}). Firstly, note that the effective Rastall mass $\tilde{M}$ depends now on $\rho$ and $p$. Also, if $\tilde{p} < 0$ the equilibrium in equation (\ref{eq:TOVR}) is lost.
Therefore, from (\ref{rhoptilde}), if the quantity $3p$ dominates the density $\rho$ then when $\alpha_2<0$ leads to $\tilde{\rho}<0$ and $\tilde{M} < 0$. 
Otherwise if the density $\rho$ dominates then $\tilde{p}<0$ and there is no equilibrium. Another aspect which we should note from the definitions (\ref{alphas}) is that $\alpha_1, \alpha_2$ and $\alpha_3$ are negative in the following interval $\eta = (-2/3,-1/2)$, $\eta = (-1/2,0)$, and $\eta = (-2,-1/2)$, respectively. 
Consequently, values $\eta<0$ may be problematic since this leads to $\alpha_2<0$. Remark that $\tilde\rho + \tilde  p = \rho + p$ and the null (and also the strong) energy condition seem to be preserved, but under extreme conditions $\alpha_2$ negative may lead to $\tilde\rho < 0$, implying violation of the energy conditions and generating instabilities. The numerical calculations to be presented in the next section confirm this possibility.

\section{Numerical Solution}\label{sec:numsol}

For solving the equilibrium equation (\ref{eq:TOVR}) we must determine the EoS of the stellar interior which, in general, has as a free parameter the central density ($\rho(r=0) = \rho_0$). For a giving central density value we obtain the corresponding mass and a radius of the star, i.e., one single point in the mass-radius diagram. One should vary ($\rho_0$) in order to obtain a curve in this plane. This is the so called mass-radius diagram.

The inner structure of the NS is still not well understood. Many groups have developed possible EoSs using microscopic many-body calculations based on, for instance, phenomenological relativistic mean-field theory and nucleon-nucleon interactions \cite{Page:2006xu, Lattimer:2012sy,Lattimer:2015eaa}. 

In order to get an insight on the internal structure of neutron stars let us firstly assume a simple pure neutron star with nucleon-nucleon interaction. In this case, a polytrope can be used as an approximation for the neutron matter equation of state
\begin{eqnarray}
 p = \kappa \rho^\gamma.
\end{eqnarray}
Following the description of an interacting Fermi gas model EoS as in Ref. \cite{Prakash} we have $\kappa \approx 2.0\times 10^5$ $\frac{cm^5}{g\;s^2}$ when $\gamma = 2$. 

\begin{figure}[!htb]
\includegraphics[width=\linewidth]{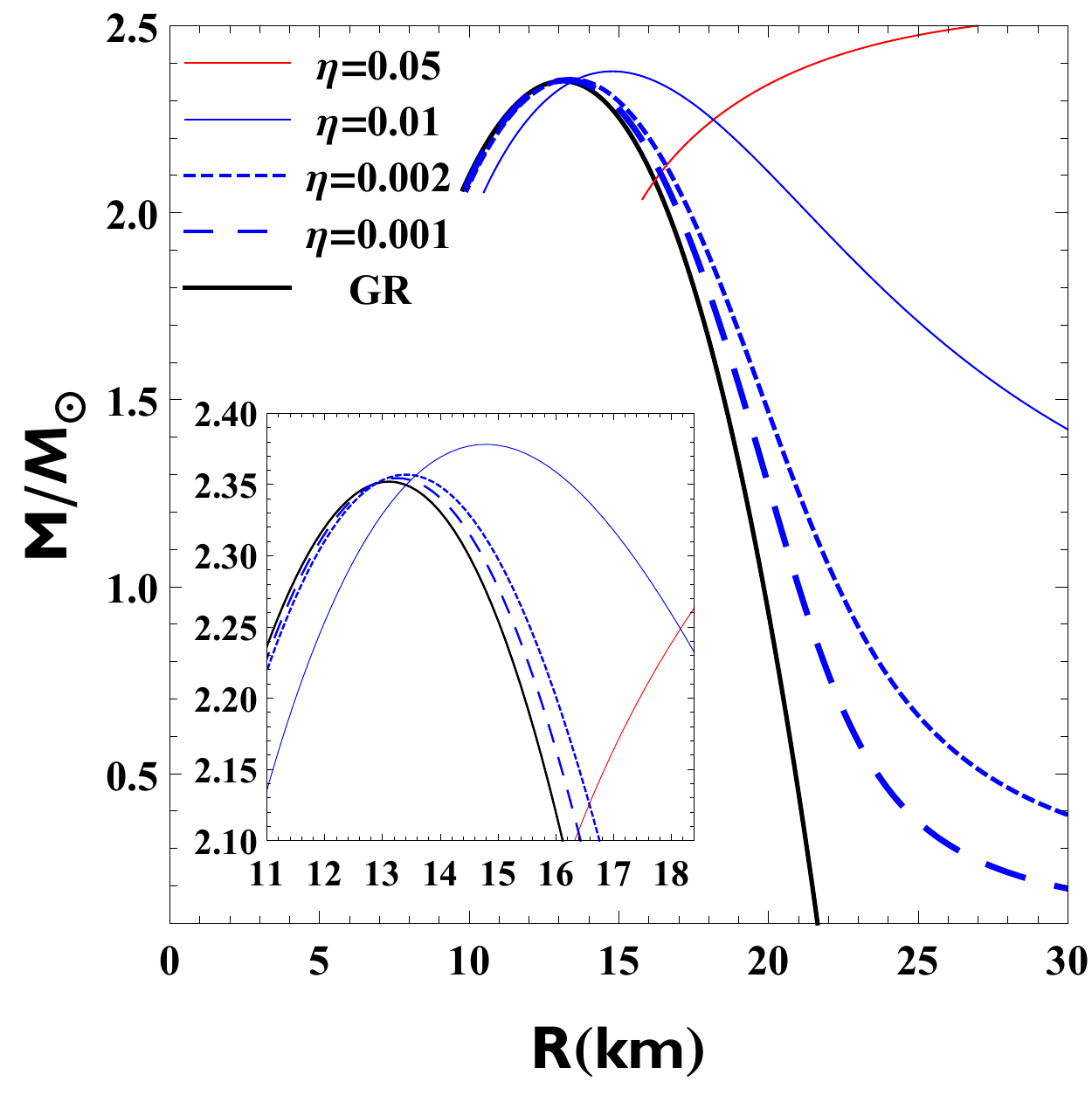}
\caption{Mass-radius contours for a pure neutron star (with interactions) using Prakash Method.}
\label{fig:Prakash}
\end{figure}

The corresponding mass-radius diagram is shown in figure (\ref{fig:Prakash}). 
Since the TOV equation is a non-trivial integro-differential equation we have used an adaptive step-size Runge-Kutta method as described in \cite{Press}.

Note that we recall in this plot the usual definition for mass as presented in (\ref{eq:massa}). The labels in the figure indicate the $\eta$ values. The black line represents the standard GR case ($\eta=0$). For $\eta = 10^{-3}$ the mass-radius profile around the maximum mass is very close to the GR. Large radii equilibrium configurations in the Rastall case are more massive than in GR. The red line ($\eta = 0.05$) clearly fails in predicting the existence of typical neutron stars with masses around $2 M_{\odot}$ within a $10-15 \quad km$ radius. It is worth noting that the numerical evaluation does not converge for $\eta < 0$. In this case the hydrostatic equilibrium equations (\ref{eq:TOVR}) are unstable and therefore no stable stellar configurations are found. This indicates a possible violation of the energy conditions.

Let us concentrate now our discussion on more realistic EoSs. The BSk family of equations aim to provide an unified description of the stellar interior treating consistently transitions between outer and inner crust (and core). They have $23$ free coefficients which have to be fitted numerically. Each BSk equation of state correspond to one specific numerical fit. 
At neutron star densities, the unified BSk19, BSk20, and BSk21 EoSs approximate, respectively, the soft, moderate, and stiff astrophysical EoSs {FPS \cite{Friedman:1981fj, Lorentz}}, APR \cite{Akmal:1998vn}, and V18 \cite{Li:2008fr} (see also \cite{Potekhin:2013si} for discussion and references).

\begin{figure}[!htb]
\includegraphics[width=\linewidth]{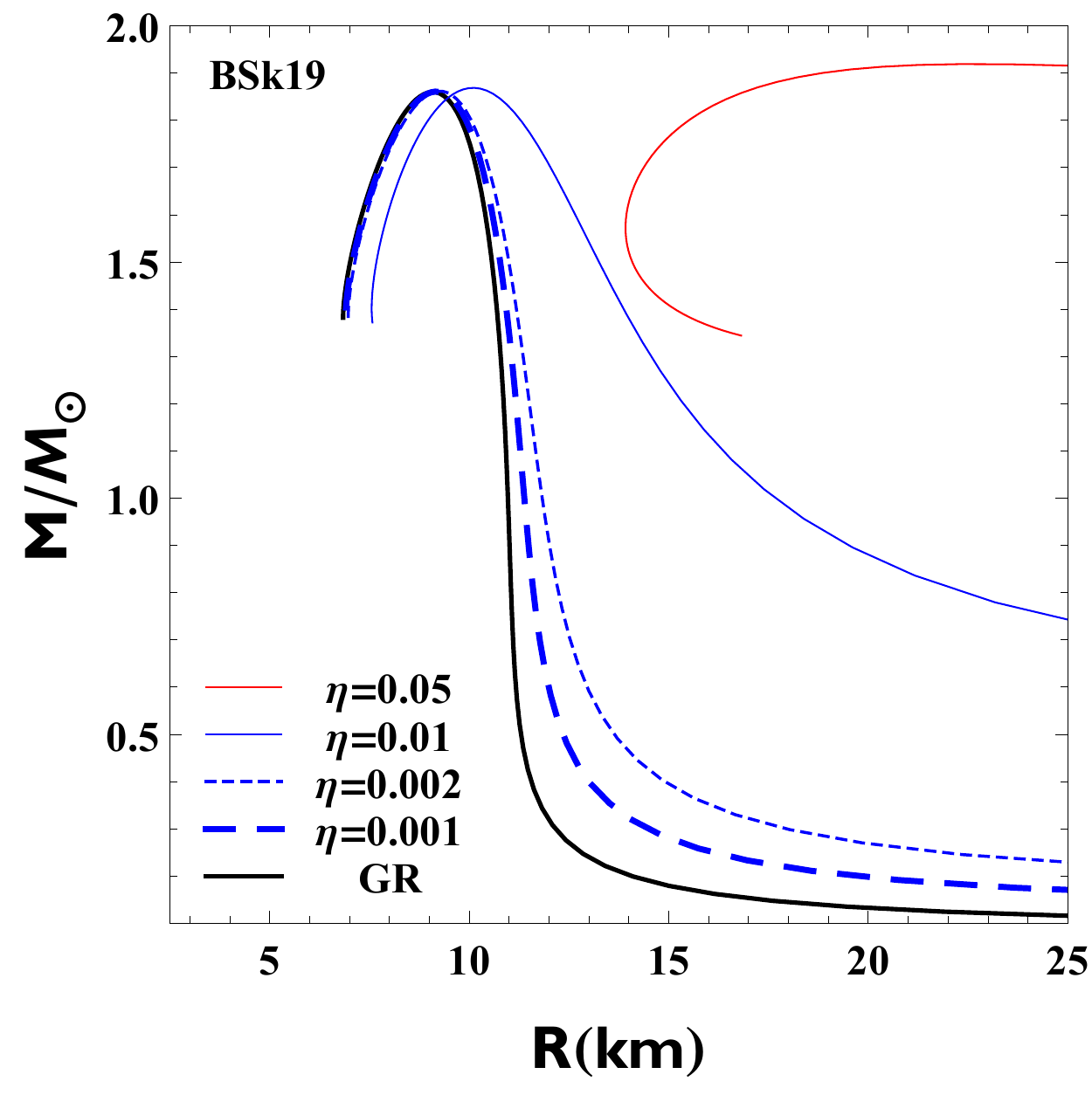}
\caption{Mass-Radius contours for representative values of $\eta$ in the BSk19 model.}
\label{fig:BSk191}
\end{figure}

\begin{figure}[!h]
\includegraphics[width=\linewidth]{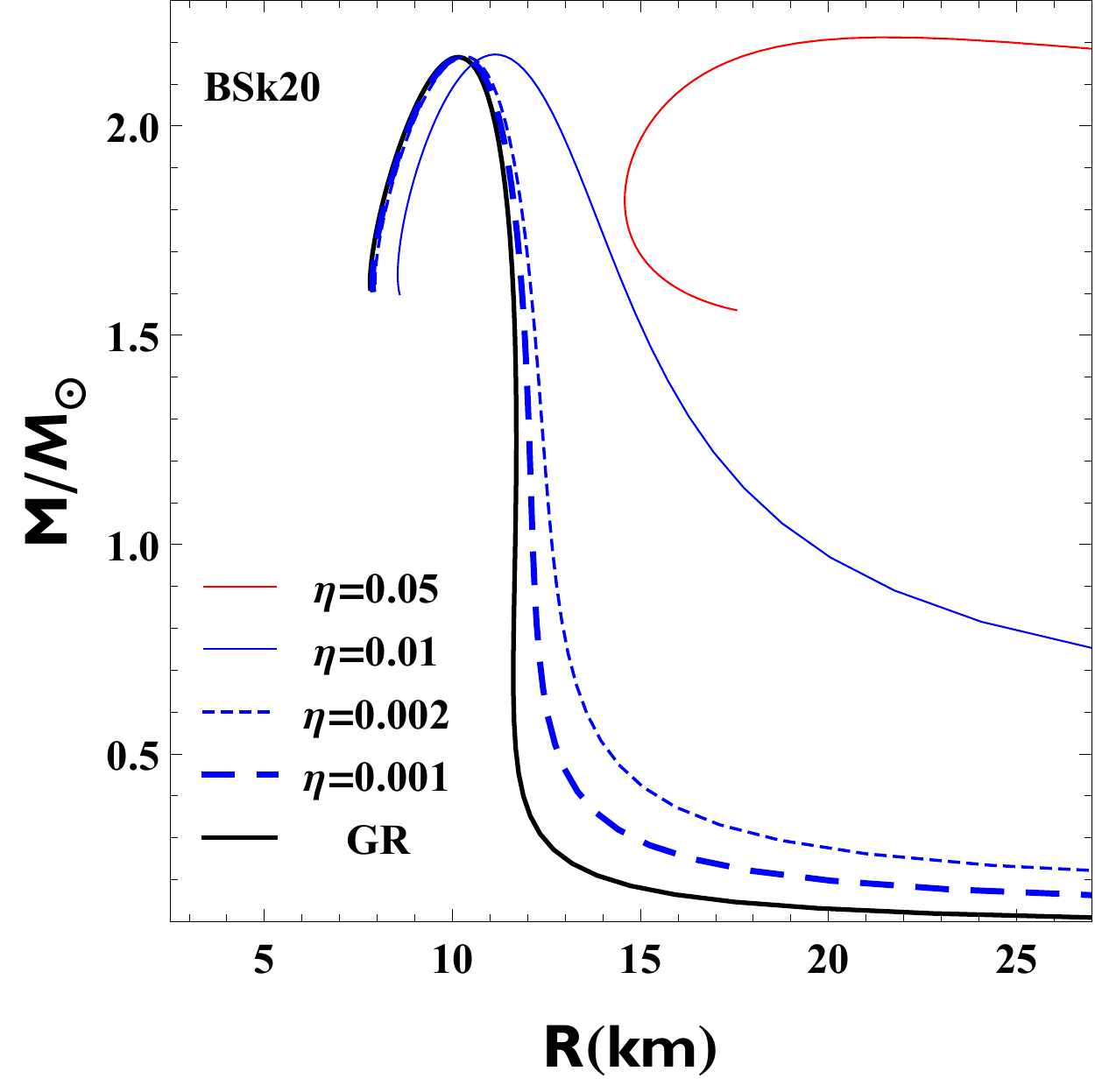}
\caption{Mass-Radius contours for representative values of $\eta$ in the BSk20 model.}
\label{fig:BSk201}
\end{figure}

\begin{figure}[!h]
\includegraphics[width=\linewidth]{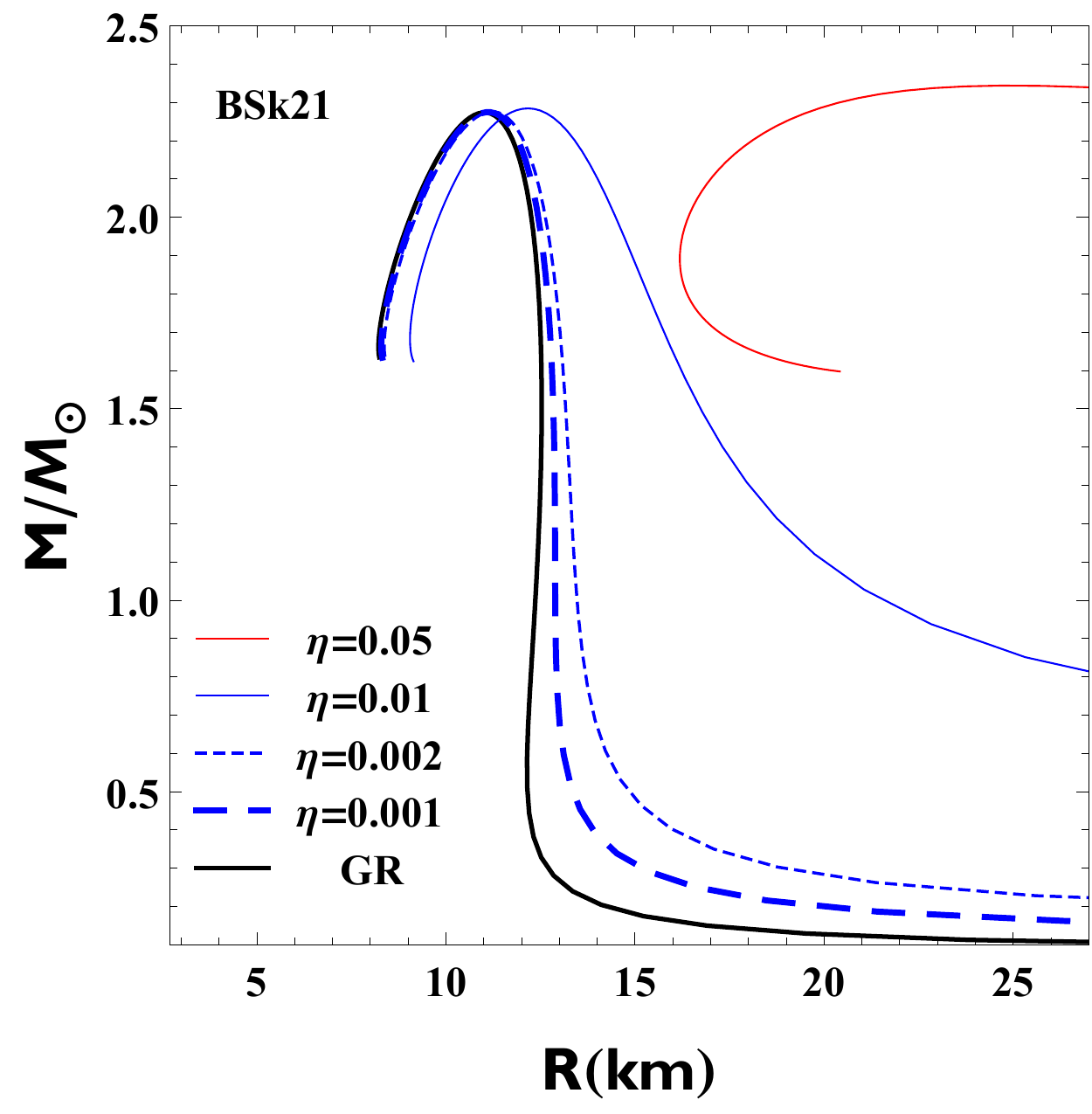}
\caption{Mass-Radius contours for representative values of $\eta$ in the BSk21 model.}
\label{fig:BSk211}
\end{figure}

Solving now the differential equation (\ref{eq:TOVR}) for these EoSs and then translating our results to the usual definition of mass (\ref{eq:massa}) as before, we find the mass-radius diagrams shown in Figs. (\ref{fig:BSk191}), (\ref{fig:BSk201}) and (\ref{fig:BSk211}), respectively.

Particularly, the BSk models have slightly distinct predictions for the maximum mass. In general, BSk19 EoS does not allow masses larger than $2 M_{\odot}$. Then, observations of very massive neutron stars with a radius $\sim 13$~$km$ would favor the BSk20 and BSk21 fits. Masses values of some observed neutron stars are usually compiled in \cite{stellarcollapse}. 

Both EoSs place a similar bound of the Rastall parameter. Clearly, $\eta=0.05$ leads to unrealistic configurations. Positive $\eta$ values showed here slightly change the maximum mass of the star. However, the most sensitive effect of changing $\eta$ values is to shift equilibrium configurations to larger radii. For example, $1.5$M$_{\odot}$ objects tipically have radii around $10 - 12$ $km$ in GR ($\eta=0$). If $\eta=0.01$ these radii values are shifted to the range $14 - 17$ $km$. Current uncertainties on the radius determination allow us to barely differentiate neutron stars radius at this level. { However, it is quite useful to quote here recent results on the determination of $R$ presented in the literature. In Ref. \cite{NSRadii}, for example, bayesian statistics has been employed to deal with the free parameters involved in modeling neutron stars. It has been found that, for the preferred equation of state, the radius of a $1.5$ M$_{\odot}$ object is $R_{1.5}=10.8^{+0.5}_{-0.4}$ $km$ and in general the peak probabilities are highly clustered in the $9 - 12$ $km$ range. Also, taking into account the results reviewed in Ref.\cite{NSR2} we can state that neutron star radii in the range $14 - 17$ $km$ should be severely disfavored. On the other hand, there are also different estimates of $R$ which are not so restricted leaving some room for admitting neutron stars with radius around $R_{1.5}\sim 15$ km \cite{suleimanov, zamfir}. Importantly, the only safe claim concerning the observed radius of neutron stars is that $R_{1.5} > 15$ $km$ are severely disfavored. The difference among the predicted values of radii occurs particularly because of the largely unknown systematic bias introduced by using the blackbody model instead of an atmosphere model or by {\it a priori} assumptions on the chemical composition of the atmosphere (see \cite{potekhin2} for a review)}. 

From the above considerations It is worth noting that giving the current uncertainties in the determination of neutron star masses and radius, values of order $\eta \lesssim 0.01$ provide stellar configurations which are consistent observations and almost indistinguishable from GR predictions.

\section{Conclusion}

In this text, we have discussed the fate of stellar equilibrium in the context of the Rastall's theory of gravity. Essentially, we have considered the case of neutron stars, where the matter conditions are extreme and the deviations of General Relativity may lead to some important effects. In fact, using three equations of state for the neutron stars, from the family of BSk models, we find that only small departures of the General Relativity are consistent.
These strong restrictions on the violation of the energy-momentum tensor conservation laws may break the degeneracy that appears at cosmological level
between Rastall's theory and the standard cosmological model based on General Relativity, the $\Lambda$CDM model \cite{cosmo}.

Since strong field effects are present in stellar systems like neutron stars, they seem to be an excellent laboratory to test deviations from the standard theories of gravity, in spite of the uncertainties existing in the study of the nuclear matter configuration in extreme density regimes.
The obtained restrictions on the Rastall's parameter (of the order of $\lesssim 1\%$ with respect to the General Relativity value, $\lambda = 1$) are, on the other hand, consistent with the interpretation of such deviations as manifestation of quantum effects in gravitational systems. {Precise future observations of $R_{1.5}$ can improve our bound to the $\lesssim 0.1\%$ level.}

Numerical studies, and considerations based on the energy conditions, restrict the Rastall's parameter to $\lambda \geq 1$. Curiously, in the study of cosmic strings in the framework of Rastall's theory carried out in reference \cite{eugenio} it was found that the space-time of the cosmic string with abelian fields becomes singular for $\lambda > 1$, i.e., $\eta >0$. However, this apparent contradiction must be faced with care. The absence of a Lagrangian in Rastall's gravity makes the full determination of the equations of motion for the matter fields challenging.
In the reference \cite{eugenio}, only the minimal extension of the General Relativity case was taken into account. To some criticisms on the Rastall's proposal based on hydrodynamics, see reference \cite{lh}.

Finally, concerning the bounds we have proposed for $\eta$ in the Rastall's theory, there is a remark to be made. The Rastall's theory leads, in the weak field limit, to a Poisson equation but for an effective gravitational coupling that is related to the usual gravitational constant $G$ by terms of order $\eta$. Measurements of the gravitational constant have a precision of order of $10^{-4}$ \cite{rosi}, but use local experiments like torsion balances or atomic interferometry \cite{quinn}.
{Even if our results may somehow be degenerated with the uncertainties on the knowledge of the gravitational coupling the uncertainty in measuring $G$ can not affect the present result at the level of $1\%$. There is also a degeneracy with
the uncertainties coming from the knowledge of the equation of state of neutron stars. Then more reliable equations of state and, especially, more precise measurements of neutron star radii are crucial to break this degeneracy.} In this sense, in the present state of art, the results obtained may indicate that Rastall's theory is still compatible with observations (in what concerns neutron stars) for a small range of values of $\lambda$, but this range can become even narrower in the future.
The results obtained in this present work reinforce that the high-density environments structures, like stars, may constraint very strongly deviations from General Relativity and, in particular, deviations from a possible violation of the usual conservation laws.

{\bf Acknowledgments:} We acknowledge enlightening conversations with Jose B. Jimenez, Oliver Piattella and Davi Rodrigues.
This work was supported by CNPq and FAPES. HV also thanks support of A*MIDEX project
(No. ANR-11-IDEX-0001-02).

\end{document}